\documentstyle[preprint,psfig]{aastex}

\def\lesssim{\mathrel{\hbox{\rlap{\hbox{\lower4pt\hbox{$\sim$}}}\hbox{$<$}}}}
\def\gtrsim{\mathrel{\hbox{\rlap{\hbox{\lower4pt\hbox{$\sim$}}}\hbox{$>$}}}}

\begin{document}

\title{Systematic Errors in the Estimation of Black Hole Masses by
Reverberation Mapping}

\author{Julian H.  Krolik} \affil{Department of Physics and Astronomy, Johns
Hopkins University, Baltimore, MD 21218}

\shorttitle{Reverberation Mass Errors}

\begin{abstract}

    The mass of the central black hole in many active galactic nuclei
has been estimated
on the basis of the assumption that the dynamics of the broad emission line
gas are dominated by the gravity of the black hole.  The most
commonly-employed method is to estimate a characteristic size-scale $r_*$
from reverberation mapping experiments and combine it with a characteristic
velocity $v_*$ taken from the line profiles; the inferred mass
is then estimated by $r_* v_*^2/G$.
We critically discuss the evidence supporting the assumption of gravitational
dynamics and find that the arguments are still inconclusive.  We then explore
the range of possible systematic error if the assumption of gravitational
dynamics is granted.  Inclination relative to a flattened system may cause a
systematic underestimate of the central mass by a factor $\sim (h/r)^2$,
where $h/r$ is the aspect ratio of the flattening.  The coupled effects of a
broad radial emissivity distribution, an unknown angular radiation pattern
of line emission, and sub-optimal sampling in the reverberation experiment
can cause additional systematic errors as large as a factor of 3 or
more in either direction.

\end{abstract}

\section{Introduction}

     The constant variability of AGN lends itself to employment as a diagnostic
of their internal structure.  Most notably, because the ionizing continuum
drives the optical/UV emission lines, the time-lag between fluctuations
in the continuum and fluctuations in the lines can be used to constrain
the distance between the two emission regions under the assumption that
it simply represents light travel-time.  This program, called ``reverberation
mapping" by analogy with the
seismic techniques used in oil exploration, has been extensively implemented,
especially in the past decade (see, for example, the reviews in Gondhalekar et
al.  1994).

   Recently there have been a number of attempts to combine
reverberation-mapping measurements of the broad emission line size scale $r_*$
with spectroscopic measurements of the line width $v_*$ in order to estimate the
mass of the central black hole (Ho 1999, Wandel et al.  1999; Kaspi et al.
2000).  These two quantities can be combined to determine the central mass if the
spread of projected velocities seen in the line width is due to motions
controlled by the gravity of the central black hole.  If that is the case, the
central mass $M \sim r_* v_*^2/G$.

     The logic of the argument depends crucially on our confidence that gravity
truly dominates the dynamics of the emission line material.  If the estimate $M
\sim r_* v_*^2/G$ is to be meaningful, we must be able to rule out significant
influence from other forces such as radiation pressure or magnetic fields.  The
quality of the evidence so far will be discussed in \S II.

     If we grant that gravity is the most important force, the next step is
to change the ``$\sim$" in the relation to an ``="; i.e., writing $M = q r_* v_*^2/G$,
we must evaluate the coefficient $q$ as accurately as possible.
In many efforts hitherto (e.g. Wandel et al.  1999, Kaspi et al.  2000),
$q$ has been implicitly taken to be the value appropriate to sources
following isotropically-oriented circular orbits at radius $r_*$.
This value is designated the ``virial" mass.  As we shall show, however,
there can be a sizable ratio between
the true mass and this ``virial" estimate.  Several factors combine to determine
this ratio.  Many of these have been mentioned and discussed qualitatively in the
literature (e.g., Wandel et al.  1999, Kaspi et al.  2000), but few have been
studied quantitatively.  A proper evaluation of the relation between the
``virial" mass and the true mass is important because the ``virial" mass is often
compared with other mass estimates (Ho 1999; Wandel 1999; Kaspi et al. 2000;
Gebhardt et al. 2000c).

     The first problem to consider is that there may be a delay between
variations in the ionizing continuum
(which is genuinely responsible for driving line emission) and the associated
variations in the continuum band that is actually observed.  Any such
delay would artificially enlarge $r_*$ by an amount equal to $c$ times the
delay.  There are some
observational indications that delays of this sort exist, but they appear (if
real) to be short compared to the continuum--line delay (Collier et al.  1998).

    Next, if all the line emission occurred within a thin spherical shell at radius
$r$, for a fixed definition of $v_*$ in terms of the line profile, the parameter
$q$ can vary over a wide range depending on the orbital shape distribution and
the orbital inclination distribution.  This source of systematic error will be
discussed in \S III.

    The detailed nature of internal (i.e., non-kinematic) properties of the
emitting matter also influence the parameter $q$, so that, in the absence of
information constraining these properties, there is an associated systematic
error (see, e.g., the qualitative discussion in Wandel et al.  1999).  The
breadth of radii across which line-emitting matter is located is one such factor.
The angular radiation pattern of the line-emitting material is another.

One way to quantify the radial emissivity distribution is through the ``transfer"
or ``response" function $\Psi$ defined by \begin{equation} \delta L_l(t) =
\int_{0}^{\infty} \, d\tau \Psi(\tau) \delta L_c(t-\tau).  \end{equation} That
is, fluctuations in the continuum luminosity $\delta L_c$ predict fluctuations in
the line luminosity $\delta L_l$ at later times through a convolution with
$\Psi(\tau)$.  Although the true relationship between line luminosity and
incident continuum flux is often nonlinear, if $\delta L_c$ and $\delta L_l$ are
interpreted as fluctuations relative to a long-term mean value, equation~1 is
valid provided the fluctuations are small compared to the mean.
At the order of magnitude level, the amplitude of $\Psi(\tau)$ tells us how much
matter there is (and how sensitive its line output is to continuum fluctuations)
at radius $r \sim c\tau$.  Although relatively few monitoring experiments
produced data of good enough quality to allow inferences of $\Psi(\tau)$ to be
made, those that have been analyzed indicate that there is significant response
across at least an order of magnitude dynamic range in lag and presumably,
therefore, across a comparable dynamic range in radius (Krolik et al.  1991; Done
\& Krolik 1996; Ulrich \& Horne 1996).

     This fact means that the characteristic $r_*$ estimated from comparing the
continuum and line light curves is a peculiar weighted average over a wide range
of radii.  Similarly, the characteristic $v_*$ is a {\it different} weighted
average of the line-of-sight velocities over that same range.  Thus, it is
unclear how to interpret the combination $r_* v_*^2$, as it mingles contributions
from different radii with different weights.

  The angular radiation pattern can introduce further factors of order unity because
gas at radius $r$ but on the near side of the continuum source to us appears to
respond more quickly than gas at the same radius but on the far side.  As
we will show in \S 4.2, the angular radiation pattern also interacts with
other sources of systematic error.

   Finally, the characteristic scales inferred from monitoring are affected
by the shape of the continuum fluctuation power spectrum.  The {\it
effective} power spectrum is determined by a combination of the intrinsic
spectrum and the details of the particular
finite sampling of the experiment because only certain timescales can be probed
by an realizable experiment.  Because this limitation amounts to a filtering in
frequency space, it effectively alters both the line and continuum fluctuation power
spectra.  Through this means, the sampling also influences the factor $q$.

     The quantitative impact of these factors (assuming regular sampling and
noise-free data) will be examined in detail in \S IV.  We do not examine the
consequences of irregular sampling (which can be expected to introduce a further
systematic error) or measurement uncertainty (which introduces a random error).
We limit ourselves in this paper to systematic errors because they can be
computed very efficiently in the frequency domain; estimating the impact of
random errors is far more efficiently done in the time domain because the error
at each point in the line profile must be evaluated.

\section{How Can We Tell That the Dynamics are Gravity-Dominated?}

   One obvious test of whether the dynamics governing the emission line region
are dominated by the gravity of a central point-mass is whether the
characteristic speed is $\propto r^{-1/2}$.  The empirical record here is mixed,
but indicates at least marginal consistency with this proposition.

    Comparing the cross-correlation peaks of different lines to the widths of
their mean profiles as measured in the IUE monitoring campaign on NGC 5548
(Clavel et al.  1991), Krolik et al.  (1991) found rough agreement with the
relation $v(r) \propto r^{-1/2}$.  Repeating this exercise with the widths of the
{\it rms} profiles (for a precise definition, see \S 4.1), Peterson \& Wandel
(1999) likewise found consistency, with somewhat smaller departures.  Peterson \&
Wandel (2000) performed the same test on two other galaxies, NGC 7469 and 3C
390.3, again finding consistency, but with poorer data.  There are only three
data points for NGC 7469 (one of which has rather large uncertainty in $v_*$) and
four for 3C 390.3 (with one having large error bars in both $r_*$ and $v_*$).  In
both cases, the dynamic range in radius is only a factor of three.  On the other
hand, Krolik \& Done (1996) found that the best fit to the velocity-resolved
monitoring data for the CIV~1549 line from the HST campaign on NGC 5548 (Korista
et al.  1995) was a rather slower dependence of velocity on radius, although the
increased $\chi^2$ created by forcing $v \propto r^{-1/2}$ was not large enough
to exclude this scaling.

     While a significant deviation from $r^{-1/2}$ scaling could rule out
point-mass gravitational dynamics, even perfect agreement could not prove it.
The reason is that several other kinds of dynamical models make the same
prediction.  These include such diverse examples as cloud outflows driven by
photoionization when the ionization parameter in the clouds is fixed (Blumenthal
\& Mathews 1975), disk winds driven by line scattering (Murray et al.  1995), and
magnetically-driven disk winds (Emmering, Blandford \& Shlosman 1992).  In fact,
at a qualitative level, both of the latter two models are in rough agreement with
the NGC 5548 velocity-resolved monitoring data (Chiang \& Murray 1996, Bottorff
et al.  1997).  In this respect, they are in as good agreement with empirical
tests as the model of point-mass gravitational dynamics.

    Peterson \& Wandel (2000) make the further argument that the mass they infer
by assuming gravitational dynamics is so large that the ratio of total luminosity
to the Eddington luminosity is too small to permit significant radiative
acceleration.  This argument is, however, flawed in two important ways.

    First, by definition, all rival models have characteristic speeds $v_*$ that
scale $\propto r^{-1/2}$.  We may therefore write $v_* = A v_{esc}$, where, as
usual, $v_{esc} = (GM/r)^{1/2}$; if the model entails outflow, $A \geq 1$.  If
that characteristic speed is then used to infer a mass {\it assuming
gravitational dynamics}, the inferred mass will be $M_{inf} = q v_*^2 r/G = qA^2
M$.  The inferred mass is then an {\it over-estimate} of the true mass when $A >
1$.  Because $M_{inf}/M \propto A^2$, the error could be substantial.

    Second, the correct criterion for driving a radiatively-accelerated wind in
emission line gas is not that $L > L_E$.  Radiation force is proportional to
opacity in optically-thin gas, and the Eddington luminosity criterion assumes
that the opacity is purely Thomson.  In fact, the opacity to ultraviolet and soft
X-ray photons presented by gas capable of radiating the observed emission lines
can easily be several orders of magnitude greater (this is, in fact, the basis of
the radiatively-driven wind models cited earlier).

     The values of $L/L_E$ inferred by Peterson \& Wandel (2000) ranged from
$\sim 10^{-3}$ to $\sim 10^{-1}$.  If $L_E$ is over-estimated by a factor of 10
(as could easily occur if there are outflows at merely three times the escape
speed), the range of $L/L_E$ inferred shifts to $\sim 10^{-2}$ -- $\sim 1$.  If
the effective opacity is $\sim 10^2$ times greater than Thomson (e.g., from H
photoionization opacity in a gas whose H neutral fraction is $10^{-4}$ facing a
continuum whose peak value of $\nu L_\nu$ occurs near 10~eV), the inferred ratio
of radiation force to gravity ranges from $\sim 1$ to $\sim 100$.  Thus, this
argument fails to provide any support for dynamics dominated by gravity.

   The only further support for broad emission line region dynamics being
dominated by the gravity of a central point-mass is the rough agreement found by
Gebhardt et al.  (2000c) and Ferrarese \& Merritt (2000b) between the central
mass inferred on the basis of the correlation with the host's bulge dispersion
(Gebhardt et al.  2000b, Ferrarese \& Merritt 2000a) and the mass inferred from
reverberation by Ho (1999) and Wandel et al.  (1999).  A similar relationship
between the central mass inferred by reverberation mapping and the narrow line
width has been found by Nelson (2000).

\section{The Orbital Shape and Inclination Distribution}

    If we grant for the sake of argument that the emission line dynamics are due
to motion in the potential of a point-mass, it remains to determine the
proportionality constant $q$.  One factor that determines this quantity is the
distribution of orbital shape and inclination.  To separate these effects from
the ones discussed in subsequent sections, suppose for the moment that the
material in question emits line radiation only when located a distance $r$ from
the central object.  Here, and in the rest of the paper, we will work in
terms of the one-dimensional line-of-sight velocity dispersion defined
by $v_*^2 \equiv \langle u^2 \rangle$, where $u$ is velocity offset in the
line profile and the average is weighted by line flux.

    In the simplest imaginable case, all the orbits might be circular (at radius
$r$) and confined to a single plane.  Then the mean-square line-of-sight velocity
is $(1/2)GM\sin^2 i/r$, where $i$ is the inclination angle of the orbital axis
relative to the line-of-sight.  The factor $q$ is then $2/\sin^2 i$.

    On the other hand, it is almost as simple to posit that the velocities are
randomly directed.  However, we do not know the orbital shapes.  Circular orbits
have $|\vec v|^2 = GM/r$, so their mean-square speed in any one direction is
$GM/(3r)$; in that case, $q = 3$.  By contrast, parabolic orbits have
$|\vec v|^2 = 2GM/r$, so that $q = 3/2$ if they are isotropic.

    More complicated models (e.g., supposing that the velocity is a sum of random
and planar components as in McLure \& Dunlop 2000) are also possible.  These, of
course, would yield a different correction factor dependent on the ratio of the
magnitudes of the two components as well as on the inclination angle.  Efforts to
infer nuclear masses from stellar kinematics have found it necessary to employ
extensive modeling in order to determine how the uncertainty in orbital shape
distribution contributes to uncertainty in the inferred mass (e.g., Gebhardt et
al.  2000a).  Similar complexities may also apply here.

     We have now found that values of $q$ predicted by equally simple models
range all the way
from a minimum of $3/2$ to a maximum (nominally) of infinity, in the event of a
thin orbital plane perpendicular to the line-of-sight.  Although a very thin
orbital plane is a bit implausible, thicknesses of $\sim 10^{-1}$ cannot yet be
easily ruled out.  If so, the range of uncertainty for $q$ from considerations of
orbital shape alone is two orders of magnitude, from $3/2$ to $\sim 200$.

     We conclude this section with a technical note.  Not all authors define
$v_*$ as the root-mean-square speed.  Some measure the FWHM of the profile
instead.  If the profile is Gaussian with dispersion $\sigma$, these quantities
are related by the expression $v_{FHWM} = 2\sqrt{2\ln2} \sigma = 2.35\sigma$
(cf. the expression
$v_{FWHM} = (2/\sqrt{3})\sigma_{3d}$ suggested by Netzer 1990 and used by Wandel
et al.  1999 and Peterson \& Wandel 2000; note that the inferred mass scales as
the square of this conversion coefficient).

\section{The Radial Emissivity Distribution}

\subsection{What Kinds of Moments are $r_*$ and $v_*$?}

    To describe the way in which the characteristic scales depend on the radial
emissivity distribution, we begin with the velocity-dependent response function
\begin{equation} \Psi(\tau,u) = \int \, dV \, {1 \over 4\pi r^2} {\partial j
\over \partial F_{ion}}\left(\vec r\right) \Phi(\hat r \cdot \hat z)
\delta\left[\tau - \tau(\vec r)\right] \int \, d^3 v \, f(\vec v, \vec r)
\delta\left(u - \vec v \cdot \hat z\right), \end{equation} where $j$ is
the local line emissivity, $\hat z$ is the direction of the line-of-sight,
$\tau(\vec r) = (r/c)(1 - \hat r \cdot \hat z)$ is the time lag corresponding
to position $\vec r$, $\Phi$
describes the angular radiation pattern of the emission (we assume that it is
oriented with respect to $\hat r$), and $f(\vec v, \vec r)$ is the velocity
distribution function at $\vec r$.  As we will discuss at greater length later in
this paper, it is desirable to define $r_*$ and $v_*$ so that they correspond, as
much as possible, to the same material.  For this reason, we will use, as
recommended by Peterson \& Wandel (1999), the variable part of the line to
measure the velocity moment as well as the characteristic lengthscale.

      Several different empirical definitions of $r_*$ have been used, but all
make use of the line-continuum total-flux cross-correlation function $C_{lc}$.
For example, some have used the peak of $C_{lc}$, whereas Wandel et al.  (1999)
and Kaspi et al.  (2000) employed the centroid of that portion of $C_{lc}$ whose
amplitude is at least 80\% of the peak amplitude.  Written in terms of $\Psi$,
the cross-correlation function is \begin{equation} C_{lc}(\tau) = {1 \over
\sigma_l \sigma_c} \int \, du \, \int \, dt \, \Psi(u) * \delta F_c \, \delta F_c
(t - \tau), \end{equation} where $\sigma_{l,c}$ are the {\it rms} fluctuations in
the line and continuum flux, $F_c(t)$ is the continuum flux at time $t$, and ``*"
denotes a convolution.  The dependence of $C_{lc}$ on the continuum fluctuation
power spectrum is more clearly displayed when it is written in terms of
Fourier-transformed quantities:  \begin{equation} C_{lc}(\tau) = \int \, du \,
\int \, df \, e^{-2\pi i f \tau} \hat\Psi |\hat{\delta F_c}|^2 , \end{equation}
where $\hat{\,}$ over a symbol (except for $\hat z$, the unit vector) indicates a
Fourier transform.  Because $\Psi(\tau)$ is a real function, it is convenient to
rewrite this expression as
\begin{equation} C_{lc}(\tau) = \int \, du \,
\int_{0}^{\infty} \, df \, 2 Re\left[e^{-2\pi i f \tau} \hat\Psi
\right]|\hat{\delta F_c}|^2 .
\end{equation}
If $\tau_*$ is defined as a centroid of the cross-correlation function, we have
\begin{equation}
\tau_* =
{\int \, d\tau \tau \int \, du \, \int_{0}^{\infty} \, df \, 2 Re\left[ e^{-2\pi
i f\tau} \hat\Psi \right] |\hat{\delta F_c}|^2 \over \int \, d\tau \int \, du \,
\int_{0}^{\infty} \, df \, 2 Re\left[ e^{-2\pi i f\tau} \hat\Psi
\right]|\hat{\delta F_c}|^2 } ,
\end{equation}
where the limits on the $\tau$ integrals are chosen according to the particular
kind of centroid desired.

     Similarly, $v_*$ may be defined in any of several ways, all based on the
{\it rms} profile.  As discussed in \S III, some use its FWHM, others its
mean-square speed.  Writing the {\it rms} profile also in terms of $\hat \Psi$
and $\hat{\delta F_c}$, we have
\begin{eqnarray}
F_{rms}(u) &=& \left[ T^{-1}
\int_{0}^{T} \, dt \, |\Psi(u) * \delta F_c|^2 \right]^{1/2} \\ &=& \left[ T^{-1}
\int \, df \, |\hat \Psi \hat{\delta F_c}|^2 \right]^{1/2} ,
\end{eqnarray}
where the second form follows from the convolution and Parseval's theorems.  If, for
example, $v_*$ is defined as the {\it rms} speed,
\begin{equation}
v_* = \left\{
{\int \, du \, u^2 \left[ T^{-1} \int \, df \, |\hat \Psi \hat{\delta F_c}|^2
\right]^{1/2} \over \int \, du \left[ T^{-1} \int \, df \, |\hat \Psi \hat{\delta
F_c}|^2 \right]^{1/2} } \right\}.
\end{equation}

    Contrasting equations 6 and 9 makes it clear that the moments over the radial
distribution resulting in $r_*$ and $v_*$ are {\it different}, so that the
characteristic scales they refer to need not coincide.  These forms also make it
obvious that the moments depend on $\hat{\delta F_c}$ in addition to the
intrinsic nature of the emission line region.

     In the preceding derivation we have tacitly assumed that an infinite amount
of data is available to determine the cross-correlation function and {\it rms}
line profile.  Real experiments do not, of course, yield infinite data trains.
However, the pairing between function and Fourier transform can be taken over
almost without alteration to discrete Fourier transforms after allowance for a
few restrictions:  the functions are assumed to be periodic with a period equal
to the duration $T$ of the data; the sampling is uniform (we call the interval
$\Delta t$); the limits on frequency in all integrals are $-1/(2\Delta t)$ to
$+1/(2\Delta t)$; and the frequency resolution is $1/T$.

    One consequence of these restrictions is that the data are effectively
filtered in such a way as to eliminate all frequencies higher than $1/(2\Delta
t)$ and lower than $1/T$.  In rough terms, there is a mapping between frequency
$f$ and distance-scale $r \sim c/f$; therefore, the character of the sampling can
bias the weighting given different distance-scales, ultimately leading to a
systematic error in the inferred mass.

\subsection{Examples}

    To explore the impact of these moments, we have used equations 6 and 9 to
evaluate $M_{inf}/M$ for a variety of choices of continuum variability behavior,
underlying physical model, and sampling.  Our goal is to separately identify the
systematic errors induced by each cause:  intrinsic character of the moments,
detailed properties of the line emission such as angular radiation pattern or
radial distribution, and poor sampling.

\subsubsection{Model definition}

     In all cases, we will use the same basic model, chosen as the simplest
non-trivial one permitting an exploration of these effects.  In this model the
line emissivity depends only on $r$, the angular radiation pattern $\Phi (\mu) =
(1 - \mu)^\gamma$ for $\gamma = 0$ or 1 (here $\mu \equiv \hat r \cdot \hat z$),
and the 1-d velocity distribution is a Gaussian with dispersion
$v(r)$.  So that effects due to orbital shape and inclination may be cleanly
separated from the other sources of systematic error, we assume in all cases
that $v(r) = (GM/3r)^{1/2}$, i.e., the orbits are circular and isotropically-oriented.
This choice implies that the correct value of $q$ is $3 M/M_{inf}$.
The response function $\Psi$ is then
\begin{equation}
\Psi (\tau,u) = \int_{c\tau/2}^{\infty} \, dr \, {c \over 2r} {\partial J \over
\partial F_{ion}}(r) {e^{-u^2/[2 v^2(r)]} \over [\pi v^2(r)]^{1/2}} \Phi\left( 1-
c\tau/r\right).
\end{equation}
Here $J \equiv \int \, d\Omega \, r^2 j$, i.e., the radial emissivity distribution.

     For the purpose of computing $\hat \Psi$, it is convenient to reverse the
order of the $r$ and $\tau$ integrations, i.e.,
\begin{equation} \hat\Psi(f,u) =
\int_{0}^{\infty} \, dr \, {c \over 2r} {\partial J \over \partial F_{ion}}(r)
{(c/r)^\gamma \over \sqrt{2\pi} v(r)} e^{-u^2/[2 v^2(r)]} \int_{0}^{2r/c} \,
d\tau \, \tau^\gamma e^{2\pi if\tau} .
\end{equation}
For any integral value of $\gamma$, the $\tau$ integral is easy to evaluate
analytically.

    It is now time to choose a physical model.  We have two goals:  exploring the
consequences of radial distributions that have a range of radial widths,
and being at least crudely consistent with what we have learned from detailed
studies of the response functions of selected AGN (Krolik et al.  1991; Horne et
al.  1991; Wanders et al.  1995; Done \& Krolik 1996; Ulrich \& Horne 1996).
Dependence on the width of the radial distribution is of special interest because
there are indications from these detailed studies that the true radial
distribution may in fact span an order of magnitude or more in radius.  Towards
that end, we write \begin{equation} {\partial J \over \partial F_{ion}} = J_o
\exp\left[- \left(\log r - \log r_o\right)^2/\left(\Delta \log r\right)^2
\right], \end{equation} where $J_o$ is an (arbitrary and irrelevant) constant,
$r_o$ is the center of the distribution of line emissivity with respect to
radius, and $\Delta \log r$ is its characteristic width in terms of $\log_{10}
r$.  We stress, however, that there are substantial uncertainties in the
emissivity distributions that have been measured, and these measurements exist
for only a handful of objects; we therefore know very little about the true range
of possibilities.  For this reason, these simulations must be regarded as purely
illustrative; real AGN may have much more complicated emissivity distributions.

    We also assume for the simulations reported here that the power density
spectrum of continuum fluctuations takes a power-law form:  $|\hat{\delta F_c}|^2
\propto f^{-n}$.  We have explored the consequences of varying $n$, and find that
in most instances it changes the results in only a minor way.  As a result of
these preliminary explorations, we decided to fix $n=1.5$ for all the calculations
reported in this paper.

   Finally, we define $r_*$ as the centroid of the portion of
the cross-correlation curve whose amplitude is greater than 80\% of the peak,
except in those cases in which the 80\% level lies beyond the range of lags where
the cross-correlation may be estimated---in those cases (which are few), we use
the cross-correlation peak to define the characteristic radius.

\subsubsection{Errors due to the moments and the underlying physics}

   We begin by defining the systematic errors due solely to the differing
moments.  These are defined by assuming an ``ideal" experiment, i.e.  one with
900 measurements spaced at an interval $\Delta t = (1/30)(r_o/c)$.  Two examples
are illustrated in Figure~1, one assuming isotropic radiation, the other assuming
$\Phi \propto 1 - \mu$.  As can be seen, in both cases the inferred mass is
biased towards values larger than the true value, but by rather more in the
anisotropic radiation case.  The reason for this distinction is clear:  When we
preferentially see regions on the far side of the center, the lag is greater than
what it would be if we could see all the emitting regions.  Thus, the estimated
$r_*$ is also greater than it should be, and the estimated mass likewise because
$M_{inf} \propto r_*$.  In both cases, the bias grows with increasing width of
the radial emissivity distribution.  The reason for this, too, is easy to see:
differing moments mean little when the underlying distribution is narrow.  For
the widest distribution we consider ($\Delta\log r = 1.5$), the error is 60\% in
the isotropic case, and a factor of 2.7 in the anisotropic case.  The method,
therefore, has an {\it intrinsic} bias toward overestimating the central mass
that can be small in favorable cases (sharply peaked radial emissivity
distributions) but considerably larger in unfavorable ones.

\begin{figure} \centerline{\psfig{file=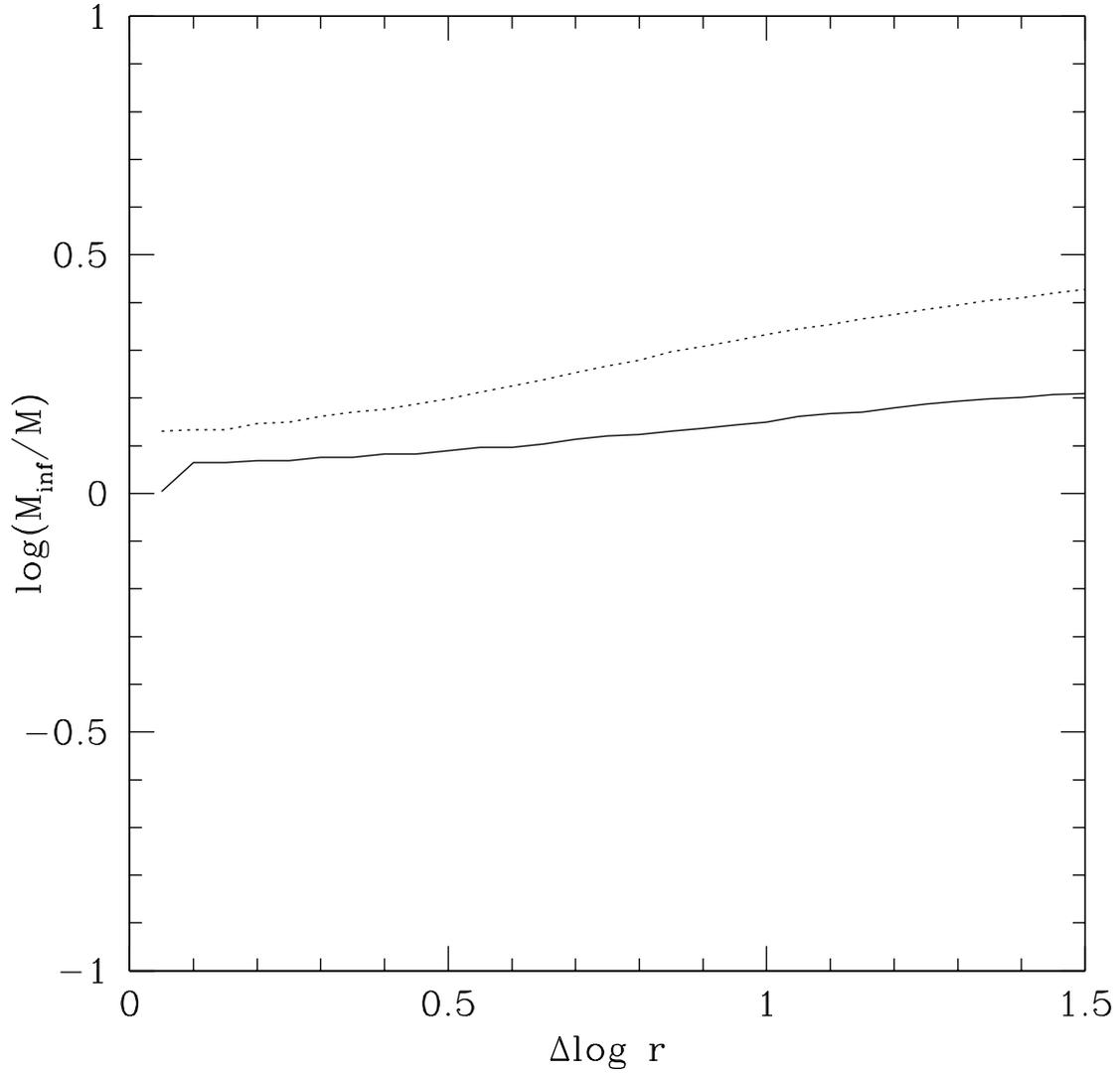,width=6.0in}} \caption{Systematic
error due solely to the differing moments that define $v_*$ and $r_*$.  The
solid curve refers to isotropic radiation, the dotted to $\Phi \propto 1 - \mu$.}
\end{figure}

\subsubsection{Errors due to sampling}

    Next we examine the effects of sampling, beginning with experiments that are
``as good as can be hoped".  We define this phrase as denoting an experiment with
64 sampling points and interval $\Delta t = (1/8)(r_o/c)$.  Judging by the
history of the experiments, this seems to be about as many measurements as can be
managed (e.g., 60 were obtained in the original IUE campaign:  Clavel et al.
1991; 39 were obtained in the HST campaign on NGC~5548:  Korista et al.  1995; of
the 28 quasars monitored by Kaspi et al.  2000, the median number of observations
was 25, although one had as many as 70).

    The expected level of systematic error from experiments ``as good as can be
hoped" is shown in Figure~2.  Contrasting that figure to Figure~1, we see that
when the emission is anisotropic in the sense chosen, the reduction in sampling
makes almost no difference at all.  Surprisingly, ``good", but less than
``ideal", sampling actually {\it decreases} the level of systematic error in the
case of isotropic radiation.  We will learn the reason for this when we study
poorer sampling.

\begin{figure} \centerline{\psfig{file=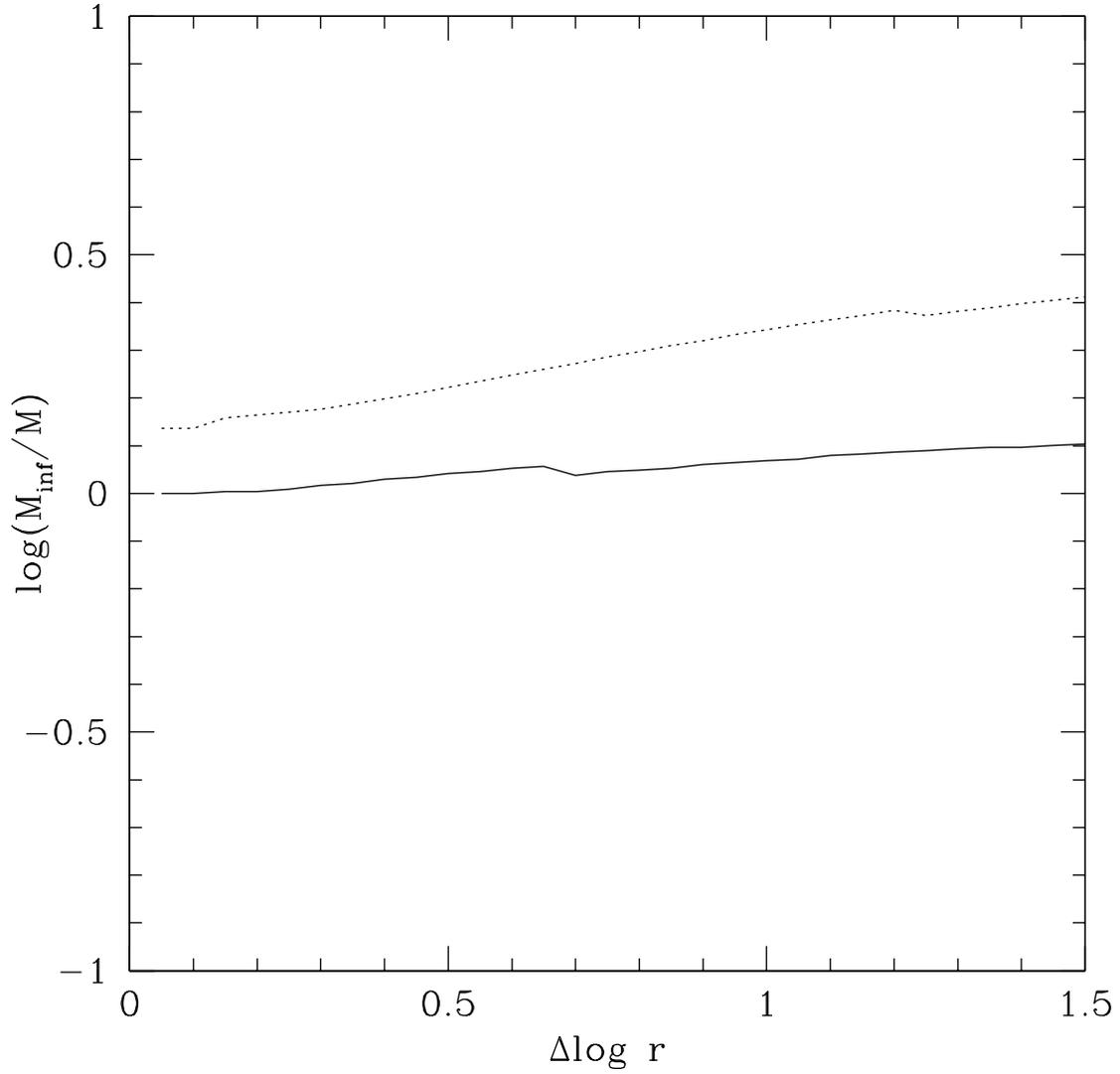,width=6.0in}} \caption{Systematic
error as a function of $\Delta\log r$ in the case of ``as good as can be hoped"
sampling.  The solid curve refers to isotropic radiation, the dotted to $\Phi
\propto 1 - \mu$.}  \end{figure}

     Truly sub-optimal sampling increases the opportunity for error.  Two kinds
of problems are possible.  The first kind is a simple shortage of data.  Instead
of having 64 measurements, as in the ``as good as can be hoped" example, there
might be many fewer.  The consequences of data sets that are too small are shown
in Figure~3.  As that figure shows, there is a systematic bias produced by short
data sets toward smaller values of $M_{inf}/M$.  When the radial distribution is
narrow, the bias is small because only a narrow range of timescales occurs in the
data.  However, as the radial distribution becomes broader, the effect grows.
Because the contrasting moments entering into $M_{inf}$ create a systematic shift
toward $M_{inf}/M > 1$ for large $\Delta\log r$, the statistical bias of short
data trains can partially counteract the systematic error induced by the moments.
It is then possible for the net systematic error to be fortuitously small.

\begin{figure} \centerline{\psfig{file=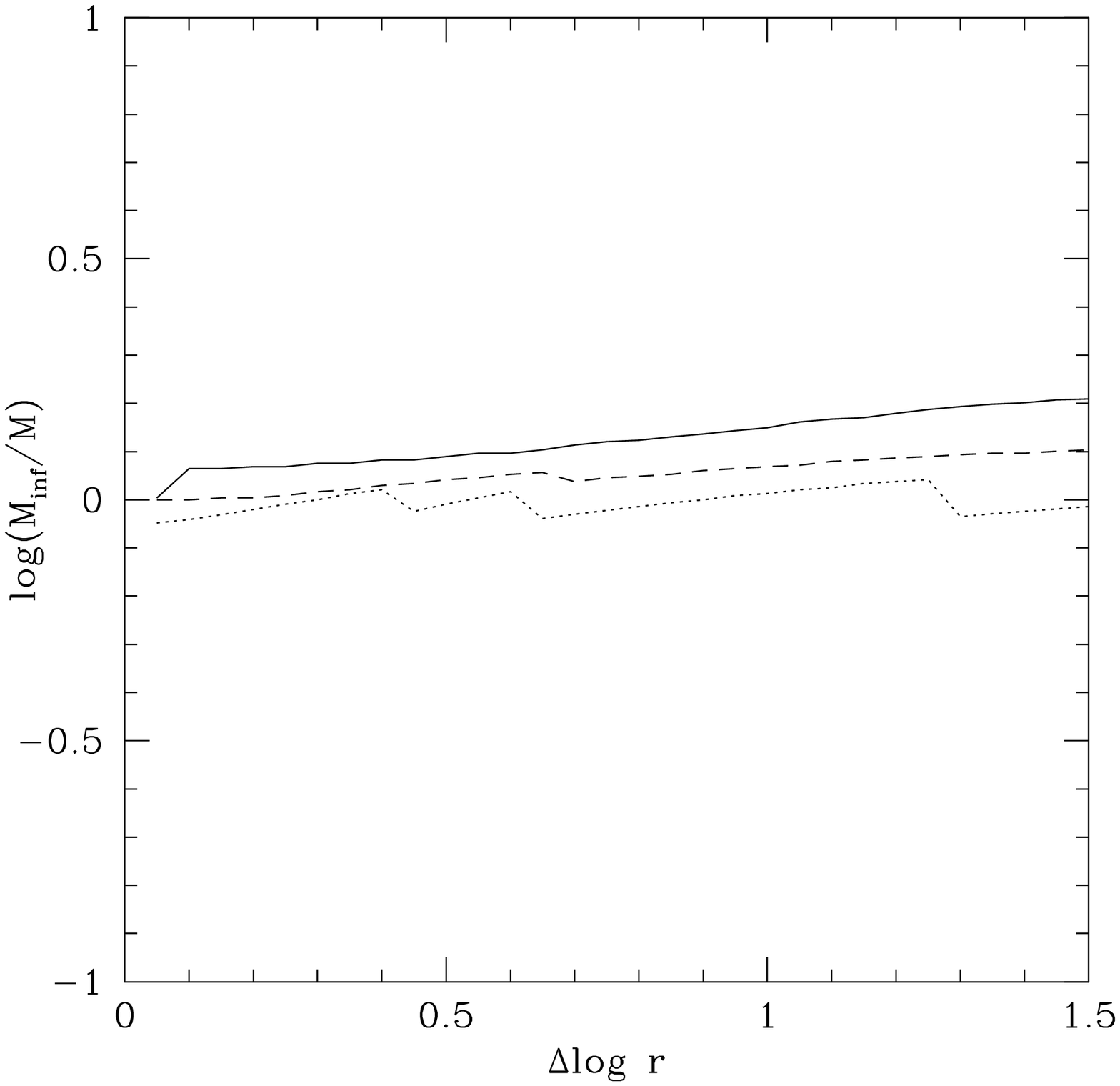,width=6.0in}}
\caption{Systematic error as a function of $\Delta\log r$ for several different
length experiments, all with timescales centered on $r_o/c$, and assuming
isotropic radiation in each case.  The solid line is the result of 900 data
points, the dashed line is the product of 64 measurements, and the dotted line
comes from a simulated experiment with only 25 points.}  \end{figure}

    The second kind of deviation from optimal sampling is an offset in timescales
sampled.  When planning a monitoring experiment, one does not know in advance
what the characteristic size of the emission line region is, although one might
estimate it by scaling from other examples (e.g., by supposing that $r_o \propto
L^{1/2}$, as might be expected on the basis of simple photoionization models; but
see the doubts raised by Kaspi et al.  2000).  Offsets in the sampling interval
relative to $r_o/c$ are therefore quite likely.  We define
$\phi = (T \Delta t)^{1/2} c /r_o$
as the offset parameter; when $\phi < 1$, the scales sampled are too small,
when $\phi > 1$, they are too large.  Figure~4 illustrates the impact
of a factor of 3 error even when a substantial data set is obtained.

\begin{figure} \centerline{\psfig{file=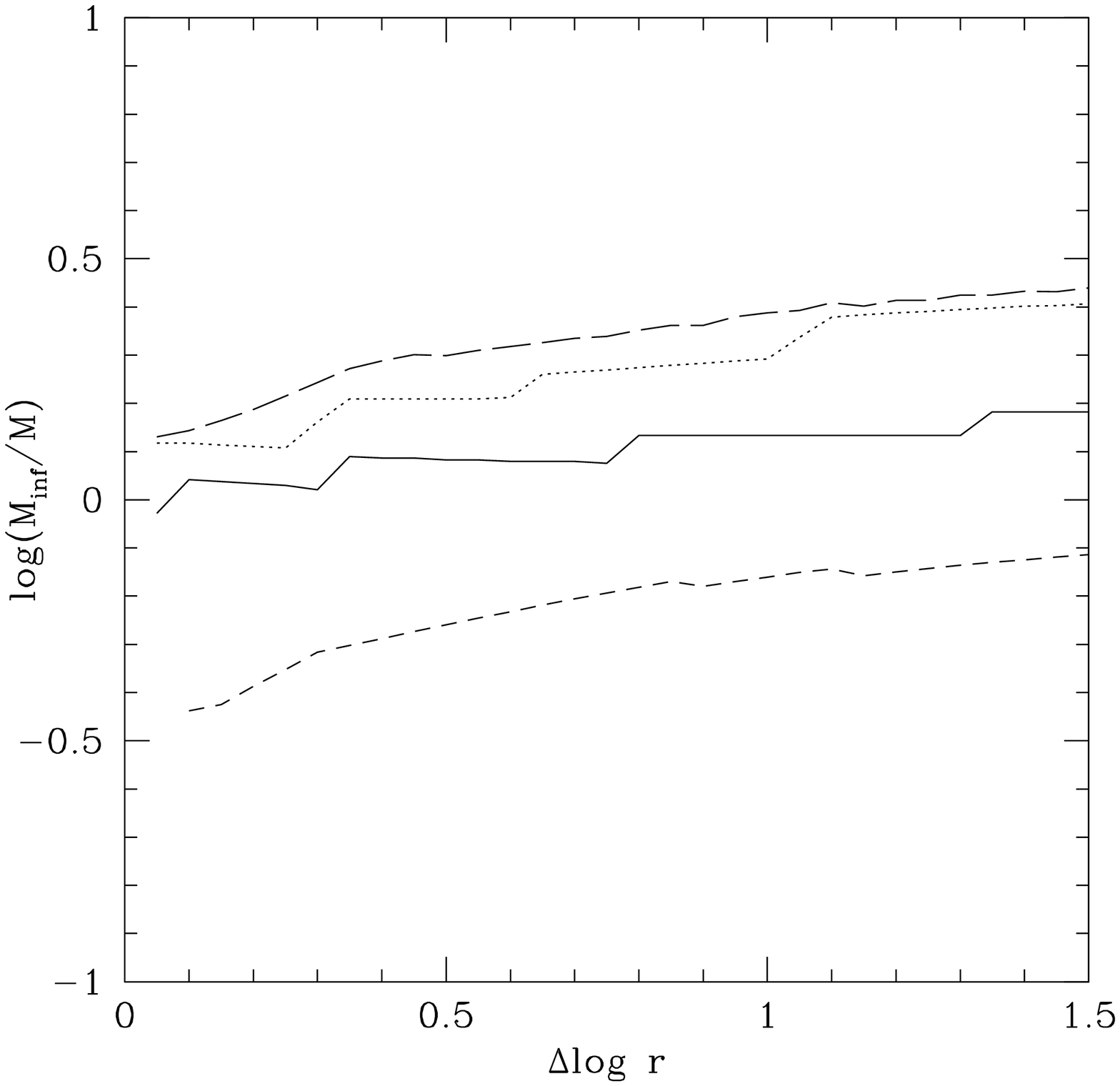,width=6.0in}} \caption{The
ratio $M_{inf}/M$ as a function of $\Delta \log r$ for four experiments with a
large number of points (64), but different sampling intervals.  The solid curve
pertains to isotropic radiation and $\Delta t = (3/8)(r_o/c)$; the dashed curve
shows the result for the same model but with $\Delta t = (1/24)(r_o/c)$.  The
dotted and long-dashed curves pertain to a model with $\Phi \propto 1 - \mu$;
$\Delta t = (3/8)(r_o/c)$ for the dotted curve, while $\Delta t = (1/24)(r_o/c)$
for the long-dashed curve.  The jagged breaks, which are particularly noticeable
in the solid curve, are due to the relatively small number of points contributing
to the cross-correlation centroid integrations.}  \end{figure}

     Larger offsets can create still larger errors.  After exploring the range
$0.01 < \phi < 100$, we find that when the radiation is isotropic,
$M_{inf}/M \simeq s(\phi,\Delta\log r)\phi$, where $s$ is a slowly-varying
function of $\phi$, such that
\begin{equation}
s(\phi,\Delta\log r) \simeq \cases{ 1 & $\phi < 1$, $\Delta\log r \ll 1$ \cr
                                    3 & $\phi < 1$, $\Delta\log r > 1$ \cr
                                  0.25 & $\phi > 1$, $\Delta\log r \ll 1$ \cr
                                  0.2 & $\phi > 1$, $\Delta\log r > 1$ \cr}
\end{equation}

  Anisotropic radiation leads to more complicated behavior.  Values of
$\phi$ between 0.1 and 10 bias the result toward $M_{inf}/M$ smaller
than unity (for $\phi < 1$) and $M_{inf}/M$ greater than unity
(for $\phi > 1$), but by smaller amounts than in the isotropic case.
More extreme values of $\phi$ push the bias sharply in
either direction, smaller values of $\phi$ leading to greatly
underestimated masses and larger values to numbers that are
substantial overestimates.  Unfortunately, unlike the case of
isotropic radiation, there is no simple approximate expression
that encapsulates these dependences.

     The origin of the sensitivity of $M_{inf}/M$ to $\phi$ lies in
the mismatch that is created when $\phi$ is grossly different from
unity between the natural timescales of the system and the
timescales probed by the experiment.  As a result, the peak in the
profile-integrated cross-correlation function can be severely
biased: for example, in the case of isotropic radiation and $\phi = 0.01$,
the characteristic scale derived from the cross-correlation centroid
is $\simeq 0.01$ times what it is when the sampling is perfectly matched
to the true scales.  That there is a peak anywhere in the measurable
range of lags is due to the existence
of some matter close to the line of sight, where it can respond quickly
despite its distance from the central source; by contrast, in the
case of $\Phi \propto 1 - \mu$, for which matter on the line of
sight is essentially invisible, the cross-correlation function is
greatest at the largest lag measurable.  Because the {\it rms}
line profile is much less strongly affected by sampling problems,
the bias in the mass estimate corresponds very nearly to the bias
in the characteristic distance scale estimate.  Not surprisingly,
tight radial distributions (i.e., small values of $\Delta\log r$)
exacerbate the timescale mismatch effect, but relatively weakly
compared to the much larger bias driven by the poor sampling.
 
\section{Discussion}

\subsection{Deciding Whether the Black Hole's Gravity Dominates Emission Line
Dynamics}

    Measuring central black hole masses by the reverberation method is possible
only if gravity dominates the dynamics of the line-emitting gas and the mass of
the black hole is much larger than any other mass within the line-emitting
region.  As was shown in \S 2, the evidence in hand to date neither proves nor
disproves this assumption.  The question naturally arises as to how this
situation might be clarified.

Point-mass gravitational dynamics might be discredited if evidence arose showing
that $v$ did not scale as $r^{-1/2}$.  One possible way to do so would be to
conduct experiments like the HST campaign on NGC 5548, but with better data, so
that $\Psi(\tau,u)$ could be more tightly constrained.  ``Better" in this context
means more epochs of observation and a combination of better signal/noise and
velocity resolution.  More epochs of observation serves both to reduce
statistical scatter and to create a larger dynamic range in timescales (and hence
length scales) probed.  Better signal/noise and velocity resolution would permit
dividing the line profile into more segments in order to achieve a greater
dynamic range in $u$.

However, demonstrating consistency with $v \propto r^{-1/2}$ is not sufficient to
prove that the black hole's gravity controls the dynamics of line-emitting gas
because there are other models that make the same prediction.  Testing these
other models vis-a-vis point-mass gravity requires independent methods.  Disk
winds might be tested by searching for the correlation between radial and
azimuthal velocity that they generically predict (Chiang \& Murray 1996).
Although this prediction is qualitatively consistent with the relatively rough
constraints posed by the HST campaign on NGC 5548, an improved experiment of the
sort described in the previous paragraph might sharpen this test.  Nonetheless,
because the kinematic predictions of so many models are so similar, providing a
direct proof of gravity-dominated dynamics will be difficult.

\subsection{Systematic Errors Granted Gravitational Dynamics:  Magnitude and
Mitigation}

   If, for the sake of argument, we grant the assumption that the motions of the
line-emitting gas are dominated by response to the black hole's gravity, both
random and systematic errors of several different varieties may cloud the result.
In this paper, we have concentrated on the {\it systematic} errors.

      One sort is due to our ignorance of the orbital shapes.  Changing from
parabolic to circular orbits alters the inferred mass by a factor of two.  Even
if the orbital eccentricity distribution is known, flattening of the emission
line region can introduce a new systematic error by eliminating our ability to
average over the velocity vectors' projections on the line-of-sight.  This error
can be as large as $\sim (r/h)^2$ for aspect ratio $h/r$.  In principle, this
latter effect could produce a very large error.

Unfortunately, whether the broad line region is round, flat, or some other shape
is still a controversial issue (Wanders et al.  1995, Dumont et al.  1998).  It
is even possible that the region responsible for some lines (e.g., the
high-ionization lines like CIV~1549) is round while the region radiating the
Balmer lines is flattened (Rokaki et al.  1992).

   Another variety of systematic error arises from the interaction between the
differing ways in which $r_*$ and $v_*$ depend on the response function, the
details of the radial emissivity distribution, the angular radiation pattern, and
the sampling.  Although in some sense each of these contributions is logically
independent, the magnitude of the combined error cannot be estimated by simply
adding them in quadrature.  For example, if the line radiation is isotropic,
relatively little systematic error is induced by the character of the moments
defining $r_*$ and $v_*$, wide emissivity distributions, or meager datasets; on
the other hand, the results obtained when the line radiation is isotropic are
very sensitive to the characteristic sampling timescale.  However, it should
also be borne in mind that all of these
conclusions were reached on the basis of exploring a very simple emissivity
model; more complicated geometries might well lead to new dependences for
the systematic errors.

   The central problem, of course, is that with only a single
monitoring dataset it is
difficult to determine which of these characteristics apply to any particular
quasar or line.  Although most photoionization models (e.g., Kwan \& Krolik 1981,
Ferland et al.  1992) predict that H recombination lines are emitted rather
anisotropically (more or less in the sense described by $\Phi \propto 1 - \mu$),
we can hardly claim to know this reliably (for an example of complications that
might change this prediction, see Kallman \& Krolik 1986).  It is possible that
other lines are also emitted anisotropically, but this suggestion is even more
model-dependent (Ferland et al.  1992, O'Brien et al.  1994).  Similarly, we do
not know anything {\it a priori} about the geometric symmetry of the
emission line region or the radial dependence of emissivity within it.  At
the same time, as the curves of Figure~2 illustrated, it is also possible for
systematic errors to cancel fortuitously.  Consequently, although we may hope
that conditions are such that the systematic error is small, it is hard at this
stage to be confident.

  A partial step forward could be provided by a two-step process applied to each
target object individually:  First, the breadth and characteristic scale of the
radial emissivity distribution can be estimated by monitoring that spans a truly
broad range of timescales; this experiment does not require good quality spectral
resolution.  Second, with that knowledge in hand, it would be
possible to design a new experiment with optimized sampling and spectral
resolution that might be better
able to control those systematic errors due to inappropriate sampling scale or
ignorance of the radial emissivity distribution.

    Another partial advance could come from making fuller use of the information
obtained in emission line monitoring experiments.  In the efforts to estimate
$M$ so far, only the moments $r_*$ and $v_*$ have been used.  There is
potentially much more information contained in the full set of time-dependent
line profiles.  If good enough data were obtained that
$\Psi(\tau,u)$ could be accurately determined, many (although not all) of
these uncertainties could be either eliminated or constrained.  For example,
as we have already remarked, the
central assumption that $v \propto r^{-1/2}$ could be tested directly.
At a more detailed level, different orbital shape and inclination
distributions can be distinguished by the contrasting shapes they give $\Psi$
in the $\tau$--$u$ plane (Welsh \& Horne 1991).  Although $\Psi(\tau,u)$
does involve an integration over radius (as in equation~10, for example),
it still provides some indication of the radial emissivity distribution.

\subsection{Random Errors}

      We stress that in this paper we have discussed only {\it systematic}
errors.  Random errors arising from flux measurement uncertainties and the
fluctuations due to specific realizations of the random processes involved add to
the final uncertainty (Peterson et al.  1998; Welsh 1999).

\subsection{Conclusions}

      Given all these considerations, it would seem that, taken in isolation,
there are systematic uncertainties in the estimation of black hole masses by
the reverberation method that are potentially large, but whose magnitudes
are difficult to estimate quantitatively.  In order to gain confidence in
the results obtained by this method, further efforts to control these
systematic errors are essential.

    Unfortunately, the underlying question of whether
gravity truly dominates the dynamics is particularly difficult to answer
securely and is rendered more difficult by the existence of uncontrolled
systematic errors.  If other, independent, measurements yield black hole masses
that coincide with those inferred from reverberation mapping (as seems possible:
Gebhardt et al.  2000c; Ferrarese \& Merritt 2000b), this agreement might
be taken as evidence in support of gravitational dynamics in the broad
line region.  Until we understand these systematic errors better, however,
it still remains possible that such agreement could be merely
the result of a fortuitous cancellation of systematic errors.

Consider one example of how this might come about:  As described in \S 2, many
models in which non-gravitational dynamics dominate predict a characteristic
speed that scales with the escape speed.  If the emission line kinematics
are interpreted as gravitational, the mass is then
consistently overestimated by a fixed ratio.  A systematic measurement error
that underestimates the mass (as would be induced by a characteristic sampling
time that is too short applied to an isotropically-radiated
emission line) could then produce an entirely
fortuitous agreement.  We would then be in the odd position of arriving
at the correct mass, but only as a result of mutually cancelling errors.

\acknowledgments

   I thank Eric Agol, Tim Heckman, and Jerry Kriss for many helpful conversations
and suggestions.  This work was partially supported by NSF Grant AST-9616922.

\end{document}